\documentclass[prl,twocolumn]{revtex4-2}
\usepackage{times}
\usepackage{amsmath}
\usepackage{amssymb}
\usepackage{hyperref}
\usepackage[normalem]{ulem}
\usepackage{setspace}
\usepackage{braket}
\usepackage{graphicx}
\usepackage{float}

\newcommand{\orcid}[1]{\href{https://orcid.org/#1}
	{\includegraphics[width=7pt]{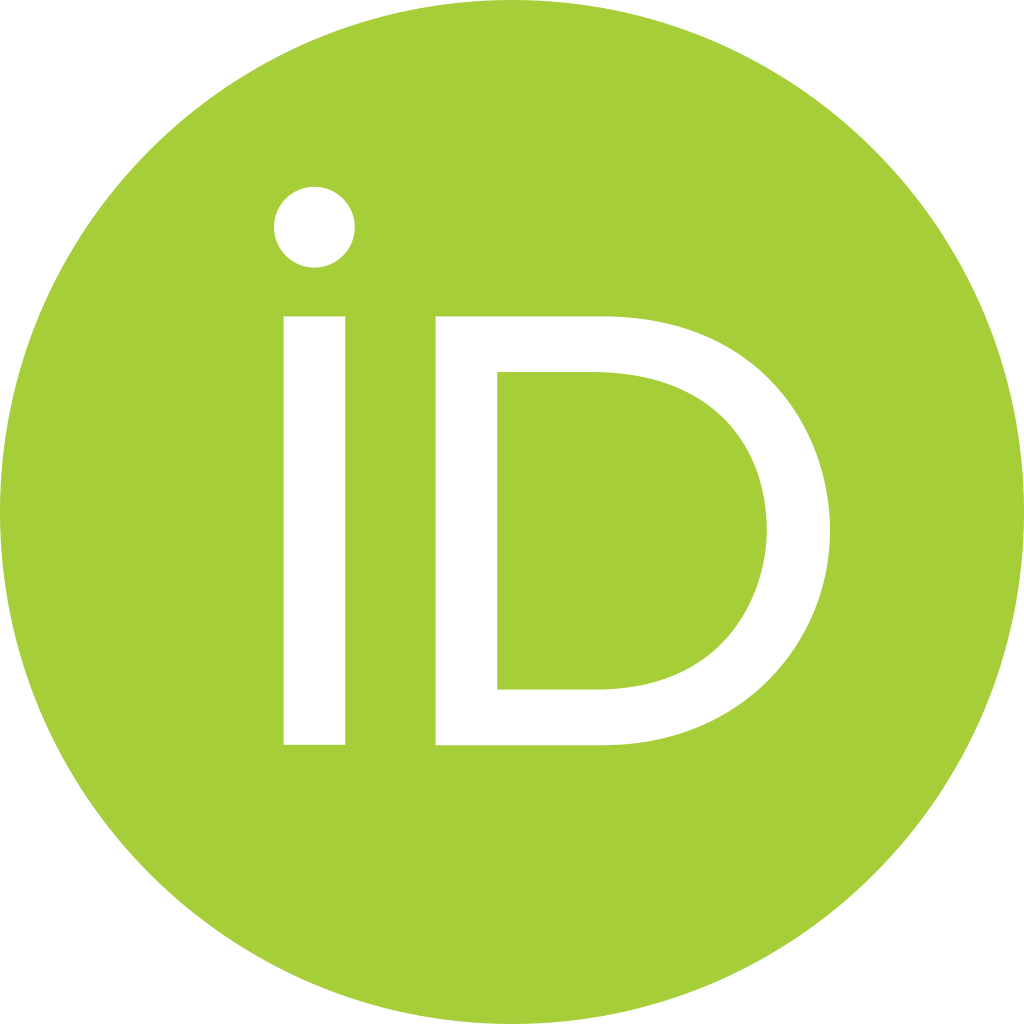}}}

\hypersetup{
    colorlinks=true,linkcolor=blue,citecolor=blue,
    filecolor=blue,urlcolor=blue,breaklinks=true
}
\RequirePackage{color}

\begin{document}

\date{\today}

\title{Rotating effects on the photoionization cross-section of a 2D quantum ring}

\author{Carlos Magno O. Pereira
\orcid{0000-0002-4009-0720}}
\email[Carlos Magno O. Pereira - ]{cmop302@gmail.com}
\affiliation{Departamento de F\'{\i}sica, Universidade Federal do Maranh\~{a}o, 65085-580 S\~{a}o Lu\'{\i}s, Maranh\~{a}o, Brazil}

\author{Frankbelson dos S. Azevedo \orcid{0000-0002-4009-0720}}
\email[Frankbelson dos S. Azevedo - ]{frfisico@gmail.com}
\affiliation{Departamento de F\'{\i}sica, Universidade Federal do Maranh\~{a}o, 65085-580 S\~{a}o Lu\'{\i}s, Maranh\~{a}o, Brazil}

\author{Lu\'{\i}s Fernando C. Pereira
\orcid{0000-0003-2452-2076}}
\email[Lu\'{\i}s Fernando C. Pereira - ]{luisfernandofisica@hotmail.com}
\affiliation{Departamento de F\'{\i}sica, Universidade Federal do Maranh\~{a}o, 65085-580 S\~{a}o Lu\'{\i}s, Maranh\~{a}o, Brazil}

\author{Edilberto O. Silva \orcid{0000-0002-0297-5747}}
\email[Edilberto O. Silva - ]{edilberto.silva@ufma.br}
\affiliation{Departamento de F\'{\i}sica, Universidade Federal do Maranh\~{a}o, 65085-580 S\~{a}o Lu\'{\i}s, Maranh\~{a}o, Brazil}

\begin{abstract}
In this letter, we investigate the nonrelativistic quantum motion of a charged particle within a rotating frame, taking into account the Aharonov-Bohm (AB) effect and a uniform magnetic field. Our analysis entails the derivation of the equation of motion and the corresponding radial equation to describe the system. Solving the resulting radial equation enables us to determine the eigenvalues and eigenfunctions, providing a clear expression for the energy levels. Furthermore, our numerical analysis highlights the substantial influence of rotation on both energy levels and optical properties. Specifically, we evaluate the photoionization cross-section (PCS) with and without the effects of rotation. To elucidate the impact of rotation on the photoionization process of the system, we present graphics that offer an appealing visualization of the intrinsic nature of the physics involved.\\

\end{abstract}

\maketitle 

\section{Introduction}

The 2023 Nobel Prize in Chemistry was awarded for the discovery and development of Quantum Dots (QDs). When particles reach sizes of just a few nanometers, the space for electrons shrinks, leading to significant effects on optical and electronic properties. One notable application of this discovery is in television screens and LED lamps. However, there are even more important potential applications under development, such as in the fields of biomedicine and environmental concerns, as well as for photon detection and photon conduction. Further details on these applications can be explored with a quick search \cite{doi:10.1021/acsanm.0c01386,doi:10.1021/acs.jpclett.7b00671,ameta2022quantum}. Motivated by the discovery of QDs, significant progress has been made in the development of experimental techniques for fabricating similar quantum structures. Notably, Quantum Rings (QRs) have been explored both experimentally \cite{PhysRevLett.84.2223} and theoretically \cite{PhysRevB.50.8460}.  
Interestingly, Tan and Inkson \cite{SST.1996.11.1635} introduced a flexible and exactly soluble model for two-dimensional semiconductors, which can describe both QDs and QRs. Subsequently, they explored the influence of the AB effect in a two-dimensional ring \cite{PhysRevB.53.6947} and investigated the magnetization and persistent currents of electrons confined within two-dimensional QDs and QRs \cite{PhysRevB.60.5626}.

In the realm of optics, it is well known that the spectrum of interband optical absorption in semiconductors is conditioned by wave functions and energy spectra of charge carriers present inside them \cite{Book.Bimberg.1999,PE.2006.31.83}. A recent study, proposed by three of us and colleagues in Ref. \cite{lima2023optical}, explores the optical absorption coefficients and refractive index changes. This study utilizes an equivalent model to the one mentioned earlier and incorporates rotating effects. On the other hand, in the study of low-dimensional semiconductors, an important focus has been on the PCS. In low-dimensional electronic systems, the photoionization process is described as an optical transition that takes place from the impurity ground state as the initial state to the conduction subbands, which requires sufficient energy in order for the transition to occur \cite{PSS.2002.232.209,JPDAP.2004.37.674}. Many studies have explored various quantum systems, considering factors such as different shapes, magnetic fields, electric fields, temperature, pressure, and their effects on PCS \cite{FAKKAHI2022115351,doi:10.1080/14786435.2021.1979267,refId0,BAGHRAMYAN2014193,EDDAHMOUNY2023140251}.
Most of these studies have involved the analysis of hydrogenic impurities. However, our primary interest lies in visualizing the impact of rotation on PCS, and thus, we do not consider hydrogenic models or any models with impurities. Actually, we are inspired by Xie's work \cite{XIE201394}, in which he investigates the effects of AB flux on PCS in a two-dimensional QR without any impurity. Since Xie's approach differs from the Tan and Inkson model, our aim is to rigorously adhere to the latter for deriving the equations that describe the system accurately.  We then make our main contribution by incorporating the effects of rotation using the Schrödinger equation with minimal coupling, as done in Ref. \cite{lima2023optical}, which uses an alternative, however corresponding approach to describe QRs.

In this letter, we present a detailed analysis of the effects of rotation on the PCS employing the Tan and Inkson model for QRs, which is widely discussed in the following. We calculate the energy states and wavefunctions, and subsequently, we conduct a numerical analysis, accompanied by graphical representations, to discuss and interpret the results. We also provide an analytical derivation for the selection rule governing the angular quantum numbers participating in the transitions, denoted by $\Delta m= \pm 1$. 

\section{Rotating quantum ring model}\label{first}

In this model, the nonrelativistic motion of a charged particle constrained to move within a localized 2D QR in a rotating frame is explored. The particle is subjected to both AB flux and a uniform magnetic field, without taking into account the spin of the particle. The Hamiltonian that describes this system is written as \cite{DANTAS201511,Book_Rotating_Frames}
\begin{equation}
H_{\Omega}=\frac{1}{2\mu}\left(\mathbf{p}-\frac{e}{c}\mathbf{A}\right)^{2}-\boldsymbol{\Omega}\cdot\mathbf{L}+ V(\mathbf{r}),\label{eq:Hr1}
\end{equation}
where \( \mu \), \(e \), \(c \), \(\mathbf{p} \), \( \mathbf{A} \), \(\boldsymbol{\Omega} \), and \(\mathbf{L}=\mathbf{r}\times \mathbf{p} \), are the effective mass,  the electric charge, light velocity, the linear momentum, the magnetic vector potential, the angular velocity, and the orbital angular momentum, respectively. 
 Also, \(V(\mathbf{r})\) is the potential of the QR, which will be defined later. Since the inherent symmetry of the system is cylindrical, we assume that $\boldsymbol{\Omega}$ has only the $z$ component, i.e., $\boldsymbol{\Omega}=\Omega \boldsymbol{\hat{z}}$, and the position vector $\mathbf{r}=\rho \boldsymbol{\hat{\rho}}$, thus $\boldsymbol{\Omega}\times\mathbf{r}=\Omega\rho\,\boldsymbol{\hat{\varphi}}$. The vector potential of the system is given in terms of the vector potential due to the solenoid (AB effect) and the vector potential of the uniform magnetic field
\begin{equation}
\mathbf{A}=\frac{B \rho}{2}\boldsymbol{\hat{\varphi}} + \frac{l\hbar}{e \rho}\boldsymbol{\hat{\varphi}},
\label{calibre}
\end{equation}
where ${B}$ is the magnetic field strength$, l=\varPhi/\varPhi_{0}$ is the AB flux parameter and $\varPhi_{0}=hc/e$ is the magnetic flux quantum.
The vector potential (\ref{calibre}) implies a superposition of magnetic fields given by
\begin{equation}
\mathbf{B}=B\,\mathbf{\hat{z}}+\frac{l\hbar}{e \rho}\delta^2(\rho)\,\mathbf{\hat{z}},\label{FieldB}
\end{equation}
where \(\delta^{2}(\rho)\) is a two-dimensional $\delta$ function, which indicates that the magnetic flux tube is located. 
It is well known in the literature that a problem described by the vector potential given by Eq. (\ref{calibre}) exhibits translational invariance along the z-direction \cite{PE.2021.132.114760}. Similarly, the eigenvalue problem in the z-direction are independent of the rotation \cite{PLA.2015.379.11,AndP.2023.535.202200371}. In addition, usually, the confinement in the z-direction is very tight so that the corresponding subband spacing is large such that we can consider that only one subband is occupied \cite{BOOK.Ihn.2010,datta1997electronic}. As a result, we can simplify the analysis by neglecting the z-direction. This allows us to analyze the electron moving in a localized disk strip that rotates with angular velocity \(\boldsymbol{\Omega}\) around the z-axis.

Now let's introduce the two-dimensional mesoscopic ring model in which the electron is confined. As mentioned earlier, this model was proposed by Tan and Inkson \cite{SST.1996.11.1635} and stands as a significant paradigm in condensed matter physics. The versatility and analytical tractability of this model continue to inspire cutting-edge research, paving the way for innovative applications and deeper insights into quantum phenomena at the mesoscopic scale. The model is described by the confining radial potential 
\begin{equation}
 V(\rho)=\frac{a_{1}}{\rho^{2}}+a_{2}\rho^{2}-V_{0}, \label{sp}
\end{equation}
where $a_{1}$ and $a_{2}$ are constant parameters, and $V_{0}=2\sqrt{a_{1}a_{2}}$. The potential (\ref{sp}) allows us to describe different types of nanostructures by a simple change of the parameters $a_{1}$ and $a_{2}$. In this model, electrons find themselves confined within a narrow ring structure, typically fashioned from semiconductor materials, allowing them to traverse the ring's perimeter \cite{PRB.1999.60.5626,AdP.2019.531.1900254,PE.2021.132.114760}. {Its profound importance lies in unveiling quantum phenomena, such as electron interference and finite-size effects, crucial in nanoscale systems \cite{fomin2013physics}.} This model's exact solvability presents a remarkable advantage, enabling comprehensive analytical insights into electron transport and magnetic properties within the ring \cite{PRB.1996.53.6947,SST.2002.17.L22,PRB.2018.98.205408}. Its precisely solvable nature grants a unique vantage point to explore fundamental quantum mechanics at a mesoscopic scale, offering deep theoretical understanding and predictive capabilities \cite{LTP.2023.49.1083}. The potential applications of this model span a wide spectrum of nanoelectronics and quantum device engineering \cite{SR.2023.13.15486,Inbook.Martins,JPCM.2011.23.115302,PE.2001.10.518}. It serves as a foundational framework for investigating and designing novel devices, such as QRs for quantum computation \cite{PLA.2019.383.1110}. 
For example, in 2D QRs, which are very small circular structures, electrons behave differently compared to macroscopic materials due to geometric constraints. Moreover, this model can be used to describe the interaction between electrons in these rings, aiding in understanding how their electronic properties are affected by the specific geometry. In QDs, which are small regions where the electron density is confined, this potential can also describe the interaction among confined electrons within that region and their quantum properties, such as discrete energy levels. In quantum wires, elongated and thin structures, this potential may be relevant for studying the behavior of electrons along the wire, considering the specific geometry of the system. In recent years, similar forms of potentials have been used to study various systems in different physical contexts, including the study of optical properties \cite{PE.2023.147.115617,JMMM.2022.556.169435,Entropy.2022.24.1059,EPJB.2015.88.83,CPL.2022.806.140000,OQE.2021.53.142,liang2011optical,xie2013optical} (see also Ref. \cite{gumber2016optical} that applies the exactly potential as in Eq. \eqref{sp} to study optical properties of a 2D QR).

At this stage, it is important to enunciate some important properties of the potential (\ref{sp}). Firstly, it has a minimum $V\left(r_{0}\right)=0$ at $r_{0}=\left(a_{1}/a_{2}\right)^{1/4}$, where $r_{0}$ is the average radius of the ring, and for $r\simeq r_{0}$, the confining potential has the parabolic form $V(r) \approx \omega_{0}^{2}\left(r-r_{0}\right)^{2}/2$, with $\omega_{0}= \sqrt{8a_{2}/\mu}$. Also, the ring width is determined by $\Delta r=\left(r_{+}-r_{-}\right)$, where $r_{-}$ and $r_{+}$ are the inner and outer radius of the QR, which are given in terms of Fermi energy by the relations \(r_{\pm}^{2}=(V_{0}+E_{F}\pm\sqrt{2E_{F}V_{0}+E_{F}^{2}})/2a_{1}\). As it is already known, depending on the chosen limits for $a_{1}$ and $a_{2}$, one can model point structures, anti-dots, and rings in two dimensions. Thus, in the limit $a_{1} \rightarrow 0$, we obtain the confining potential for a QD. In turn, for $a_{2} \rightarrow 0$, we have an anti-dot potential. Furthermore, in the general case, where $a_{1}$ and $a_{2}$ are non-zero, we have a harmonic confining potential model for a QR. 

Considering the fields and potentials defined above, the Schrödinger equation to be solved is
\begin{align}
  &\frac{1}{2\mu}\left(\mathbf{p}-\frac{e}{c}\mathbf{A}-\mu\mathbf{A}_{\Omega}\right)^{2}\psi-\frac{1}{2}\mu\mathbf{A}_{\Omega}^{2}\psi +\frac{a_{1}}{\rho^{2}}\psi+a_{2}\rho^{2}\psi \notag \\ &=(E+V_{0})\psi,\label{sh}
\end{align}
where $\mathbf{A}_{\Omega}=\boldsymbol{\Omega}\times\mathbf{r}$ can be understood as a gauge field of the rotating frame. We use eigenfunctions in the form
\begin{equation}
\psi(\rho,\varphi)=e^{im\varphi}R(\rho),\label{ansatzr}
\end{equation}
where $m=0,\pm1,\pm2,\pm3,\cdots$ is the angular momentum quantum number. After replacing (\ref{ansatzr}) in (\ref{sh}), we arrive at the radial differential equation 
\begin{equation}
R^{\prime\prime}+\frac{1}{\rho}R^{\prime}+\left(-\frac{L^{2}}{\rho^{2}}-\frac{\mu^{2}}{4\hbar^{2}}\varpi^{2}\rho^{2}+\gamma^{\prime}\right)R =0,\label{eq:radialrr}
\end{equation}
where we have defined the parameters
$\omega_{c}=eB/\mu c$, $\varpi^{2}=\omega_{c}^{2}+4\Omega\omega_{c}+\omega_{0}^{2}$, $L^{2}=\left(m-l\right)^{2}+2\mu a_{1}/\hbar^{2}$, and $\gamma^{\prime}=\mu\omega^{*}\left(m-l\right)/\hbar+2\mu\left(V_{0}+\mathcal{E}\right)/\hbar^{2}$, with $\omega^{*}=\omega_{c}+2\Omega$. Defining the new variable, $\xi=\mu\varpi\rho^{2}/2\hbar$,
we can rewrite Eq. (\ref{eq:radialrr}) as 
\begin{equation}
\xi\frac{d^{2}R}{d\rho^{2}}+\frac{dR}{d\rho}+\left(-\frac{L^{2}}{4\xi}-\frac{\xi}{4}+\gamma\right)R=0,\label{eq:rn}
\end{equation}
where \(\gamma=\hbar\left[2\mu(V_{0}+\mathcal{E})/\hbar^{2}+\mu\omega^{*}\left(m-l\right)/\hbar\right]/2\mu\varpi \). Performing an asymptotic analysis of the Eq. (\ref{eq:rn}) for the limits
$\xi\rightarrow0$ and $\xi\rightarrow\infty$, we see that the radial solution has the form 
\begin{equation}
R(\xi)=e^{-\frac{\xi}{2}}\xi^{\frac{|L|}{2}}\zeta(\xi),\label{eq:ansatzzeta}
\end{equation}
where $\zeta(\xi)$ is a function to be determined that satisfies the following differential equation:
\begin{equation}
\xi \frac{d^{2}{\zeta}}{d \rho^{2}}+\left(|L|+1-\xi\right)\frac{d\zeta}{d \rho}-\left(\frac{|L|+1}{2}-\gamma\right)\zeta=0.\label{hp}
\end{equation}
Equation (\ref{hp}) is a differential equation of the Kummer type and its solution is given in terms of the confluent hypergeometric function. Therefore, it can be shown that the  energy eigenvalues and eigenfunctions of the problem are
\begin{align}
&\psi_{n,m}\left(\rho ,\varphi \right)  =\frac{1}{\sqrt{2\pi}\lambda_{0}}
\left[ \frac{\Gamma \left(n+L+1\right)}{n!\left( \Gamma \left( L+1\right)
\right)^{2}}\right]^{\frac{1}{2}}  \nonumber \\
& \times e^{-im\varphi }e^{-\frac{\rho^{2}}{4\lambda_{0} ^{2}}}\left( \frac{\rho ^{2}}{2\lambda_{0} ^{2}}\right) ^{\frac{L}{2}}\, _{1}F_{1}\left( -n,L+1,\frac{\rho^{2}
}{2\lambda_{0} ^{2}}\right),
\end{align}
where $\lambda_{0}=\sqrt{\hbar/\mu \varpi}$, and
\begin{align}
&\mathcal{E}_{nm} =\left(n+\frac{1}{2}\sqrt{\left(m-l\right)^{2}+\frac{2\mu a_{1}}{\hbar^{2}}}+\frac{1}{2}\right) \notag \\& \times\hbar\sqrt{\omega_{c}^{2}+4\Omega\omega_{c}+\omega_{0}^{2}} -\frac{\hbar}{2}\left(m-l\right)\left(\omega_{c}+2\Omega\right)-\frac{\mu}{4}\omega_{0}^{2}r_{0}^{2}.
\end{align}
As expected, in the absence of rotation, we obtain the same result as in Ref. \cite{PRB.1999.60.5626}. Later, we will apply these solutions to analyze how rotation affects the PCS of the system.

\section{Photoionization process and selection rule}
\label{photo}

The photoionization process signifies an optical transition originating from the ground state to the continuum of conduction subbands, which commences beyond the confining potential of the QD.
The PCS represents the probability that a bound electron can be released by suitable radiation with energy $\hbar \omega$ of a specific frequency. Its magnitude is profoundly influenced by the confinement potential and photon energy \cite{articleTshipa,tshipa2021photoionization}. 

The dependence of excitation energy in the photoionization process derived from Fermi's golden rule using the well-known dipole approximation can be calculated as \cite{lax1954,PRB.2008.77.045317}
\begin{align}
\sigma \left( \hbar \omega \right) =C_{n_{r}}\hbar \omega 
\sum\limits_{f} &\left\vert \left\langle \psi _{n,m}^{(i)}\left\vert \mathbf{r}%
\right\vert \psi_{n^{\prime },m^{\prime }}^{(f)}\right\rangle \right\vert
^{2}\notag \\ &\times \delta \left( \mathcal{E}_{n,m}^{\left( f\right) }-\mathcal{E}_{n,m}^{\left( i\right)
}-\hbar \omega \right),\label{cs}
\end{align}
where 
\begin{equation}
C_{n_{r}}=\left(\frac{\xi_{\text{eff}}}{\xi_{0}}\right)^{2}\frac{n_{r}}{\epsilon}\frac{4\pi^{2}}{3}\alpha_{\text{fs}},
\end{equation}
and the function \( \delta \left(\mathcal{E}_{n,m}^{\left( f\right)}-\mathcal{E}_{n,m}^{\left( i\right) }-\hbar
\omega \right) \) is given 
by a narrow Lorentzian,
\begin{equation}
\frac{1}{\pi}\frac{\hbar \Gamma_{f}}{\left(
\mathcal{E}_{n,m}^{\left(f\right)}-\mathcal{E}_{n,m}^{\left( i\right)}-\hbar \omega \right)
^{2}+\left(\hbar \Gamma_{f}\right)^{2}}.   
\end{equation}
In the above equations, $\mathbf{r} = \rho \cos \varphi$, $\hbar \omega$ is the energy of the photon, \(n_{r}\) denotes the refractive index of the semiconductor, 
$\alpha_{\text{fs}}=e^{2}/\hbar c$ represents the fine structure constant, $\epsilon$ represents the medium's dielectric constant, and $\Gamma_{\text{f}}$ represents the relaxation rate of the initial and the final states. The quantity \(\xi_{\text{eff}}/\xi_{0}\) represents the ratio of the effective electric field \(\xi_{\text{eff}}\) of the incident photon to the average field \(\xi_{0}\) in the medium.
Let's define \(\langle \psi_{n,m}^{(i)} |{\bf{r}} |\psi_{n,m}^{(f)} \rangle \equiv \mathcal{M}_{if}\), which indicates the matrix element between the initial and final states of the dipole moment. The functions \(\psi_{n,m}^{(i)}\) and \(\psi_{n,m}^{(f)}\) symbolize the wave functions of the initial and final states, while \(\mathcal{E}_{n,m}^{(f)}\) and \(\mathcal{E}_{n,m}^{(i)}\) correspond to the respective energy eigenvalues of the transition. The transition matrix element is calculated explicitly as
\begin{align}
&\mathcal{M}_{if} =\frac{1}{2\pi \lambda_{0}^{2}}\left[ \frac{\Gamma \left(
n+M+1\right) }{n!\left( \Gamma \left(M+1\right) \right) ^{2}}\right]^{\frac{1}{2}}\left[ \frac{\Gamma \left(n^{\prime }+M^{\prime }+1\right)}{n^{\prime}!\left( \Gamma \left(M^{\prime}+1\right) \right) ^{2}}\right] ^{\frac{1}{2}}  \notag \\
& \times \int_{0}^{\infty }\int_{0}^{2\pi }e^{i\left( m^{\prime }-m\right)
\varphi }F\left( -n,M+1,\frac{r^{2}}{2\lambda _{0}^{2}}\right) \cos \varphi 
\notag \\
& \times e^{-\frac{r^{2}}{2\lambda_{0}^{2}}}\left( \frac{r^{2}}{2\lambda_{0}^{2}}\right) ^{\zeta }F\left( -n^{\prime },M^{\prime}+1,\frac{r^{2}}{2\lambda _{0}^{2}}\right) r^{2}drd\varphi,\label{mif}
\end{align}
with $\zeta =\left( M+M^{\prime }\right) /2$.
Defining the new variable $z=r^{2}/2\lambda _{0}^{2}$, Eq. (\ref{mif}) is written as
\begin{align}
\mathcal{M}_{if}& =c_{nm}^{\prime }\int_{0}^{2\pi }e^{i\left( m^{\prime
}-m\right) \varphi }\cos \varphi d\varphi \int_{0}^{\infty }e^{-z}z^{\zeta +
\frac{1}{2}}  \notag \\
& \times  \,_{1}F_{1}\left( -n^{\prime},M^{\prime }+1,z\right)
\,_{1}F_{1}\left( -n,M+1,z\right) dz,
\end{align}
where
\begin{equation}
c_{nm}^{\prime }=\frac{\lambda _{0}}{\sqrt{2}}\left[ \frac{\Gamma \left(
n^{\prime }+M^{\prime }+1\right) }{n^{\prime }!\left( \Gamma \left(
M^{\prime }+1\right) \right) ^{2}}\right] ^{\frac{1}{2}}\left[ \frac{\Gamma
\left( n+M+1\right) }{n!\left( \Gamma \left( M+1\right) \right) ^{2}}\right]
^{\frac{1}{2}}.
\end{equation}%
Using the identity $2\cos \varphi =e^{i\varphi }+e^{-i\varphi }$, the integral in the coordinate $\varphi $ results
\begin{equation}
\int_{0}^{2\pi }e^{i\left(m^{\prime}-m\right) \varphi }\cos \varphi
d\varphi =\pi \left( \delta_{m^{\prime },m-1}+\delta_{m^{\prime},m+1}\right).\label{ip}
\end{equation}
With this result (Eq. (\ref{ip})), the matrix element $\mathcal{M}_{if}$ is given by
\begin{align}
&\left\vert \left\langle \psi_{n,m}\left\vert \mathbf{r}\right\vert \psi_{n^{\prime },m^{\prime }}\right\rangle \right\vert =c_{nm}^{\prime
}\left( \delta _{m^{\prime },m-1}+\delta_{m^{\prime },m+1}\right)
\int_{0}^{2\pi }dz e^{-z}  \notag \\
& \times z^{\zeta +\frac{1}{2}}\,_{1}F_{1}\left( -n^{\prime },M^{\prime
}+1,z\right)\,_{1}F_{1}\left( -n,M+1,z\right) dz.
\end{align}
From this result, we see that non-null matrix elements follow the selection rule $\Delta m=\pm 1$. 

\subsection{Numerical results and discussions}
\label{results}

In this paper, our main focus is to understand the influence of the rotation parameter on the PCS of a rotating QR. We conduct a numerical investigation, exploiting relevant data readily available in the literature, particularly for the case of a 2D GaAs QR. From now on, we will delve into a detailed discussion of these results, derived through the application of the mathematical expressions discussed earlier.

All calculations were performed using the following physical parameters: \(\hbar\Gamma_{f}=0.1\, \text{meV}\) \cite{EPJB.2021.94.129}; \(\epsilon=13.1\); \(\mu = 0.067 \,\mu_e\), where \(\mu_e=9.1094\times10^{-31}\,eV/c^{2}\); \(\epsilon_0=8.854\times10^{-12}\,\text{F/m}\); \(\mu_0=4\pi\times10^{-7}\,\text{Tm/A}\); and \(c=2.99\times10^{8}\,m/s\) \cite{PM.2019.99.2457}. We have also considered the particular values: \(n_{r}=3.15\);  \(\alpha_{\text{fs}}=1/137\); and \({\xi_{\text{eff}}}/{\xi_{0}}=1\) \cite{PBB.2008.77.045317}. 
The rings used in the analysis have radii $r_{0}=12$ nm, $r_{0}=14$ nm and $r_{0}=16$ nm. The confinement energy is the same in all three cases and is given by $\hbar \omega_{0}=25$ meV.
In the following, we evaluate two state transitions: the first from (${n=0,m=0}$) to (${n=0,m=-1}$) and the second from (${n=0,m=0}$) to (${n=0,m=1}$). While the magnetic field is a significant factor in the eigenfunction and energy expressions, its contribution is numerically small compared to other terms. It does not play a crucial role in the physical characteristics of the system. Therefore, in the following analysis, we will set its value to $B=1\,\text{Tesla}$. 

\begin{figure}[!b]
    \centering
    \includegraphics[scale=0.55]{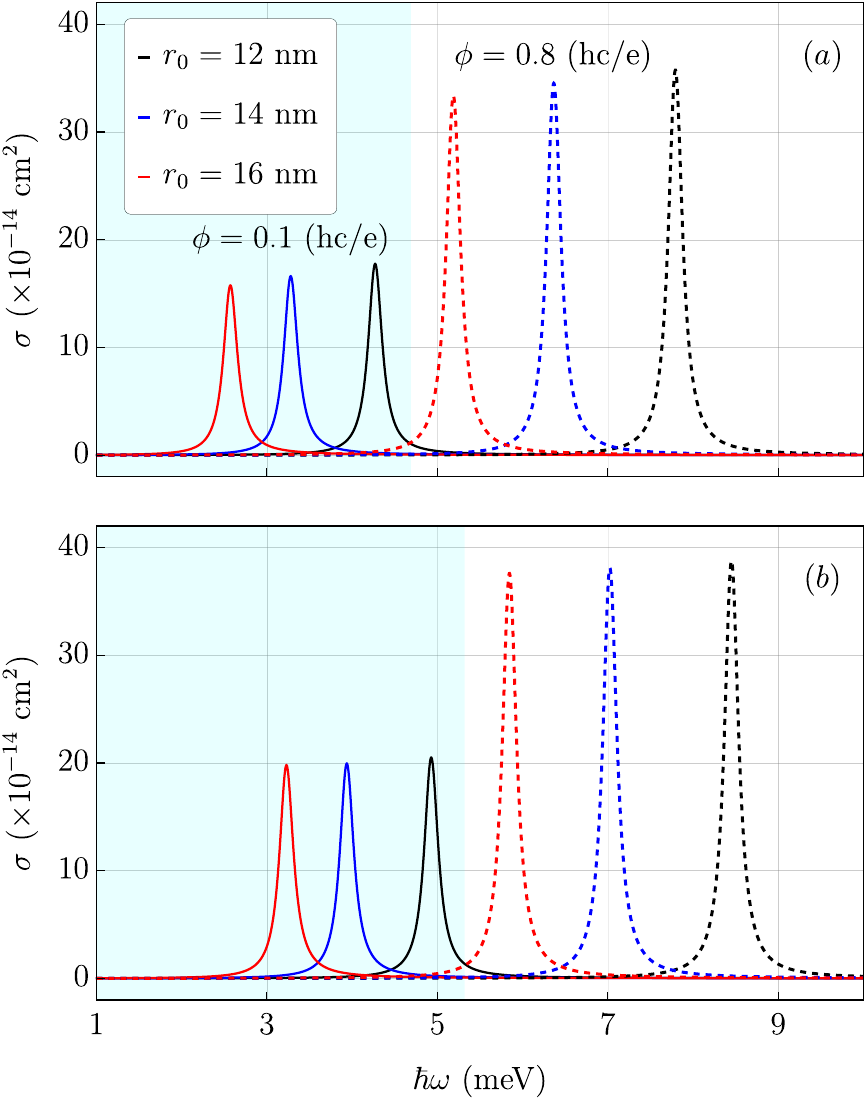}
    \caption{Graphs illustrating the PCS as a function of photon energy for various values of the average radius $r_{0}$, with a fixed value of $\hbar\omega_{0}=25\, \text{meV}$. The graphs specifically depict the transition (${n=0, m=0}$) to (${n=0, m=-1}$), where (a) corresponds to $\Omega= 0$ and (b) corresponds to $\Omega= 1\, \text{THz}$. The bluish region with solid curves corresponds to a fixed value of the magnetic flux, $\phi=0.1\, (hc/e)$, and the dashed curves outside this region represent the value of magnetic flux $\phi=0.8\, (hc/e)$.} 
    \label{fig:flux}
\end{figure}

In Fig. \ref{fig:flux}, we evaluate the transition (${n=0, m=0}$) to (${n=0, m=-1}$). We specifically highlight the influence of the AB flux on the PCS as a function of the energy of the incident photon while varying the average radius $r_{0}$ of the quantum ring and including rotation. Notably, for a fixed $\phi$, a decrease in $r_{0}$ corresponds to an increase in the amplitude of the PCS peaks. Additionally, there is a deviation in peak values, moving away from the origin along the horizontal axis as $r_{0}$ decreases.
A slight variation in $\phi$ results in a noticeable jump in the peak values of PCS amplitude and a shift in the peak position along the horizontal axis. The peak position undergoes a shift towards higher energies with an increase in magnetic flux.
 Rotation also plays a crucial role, with the PCS amplitude increasing when rotation is considered: the peaks in Fig. \ref{fig:flux} (a) are shorter than those in Fig. \ref{fig:flux} (b). 
 These observations emphasize the significant influence of magnetic flux and rotation on the probability of the optical transition of the quantum ring's state. As expected, the position of the peak in the PCS along the horizontal axis is determined by the energy difference associated with the evaluated state transition. Furthermore, the graphs indicate a substantial amplitude of the PCS, coupled with a broad range of photon frequencies that facilitate the photoionization process.
\begin{figure}[!t]
    \centering
    \includegraphics[scale=0.55]{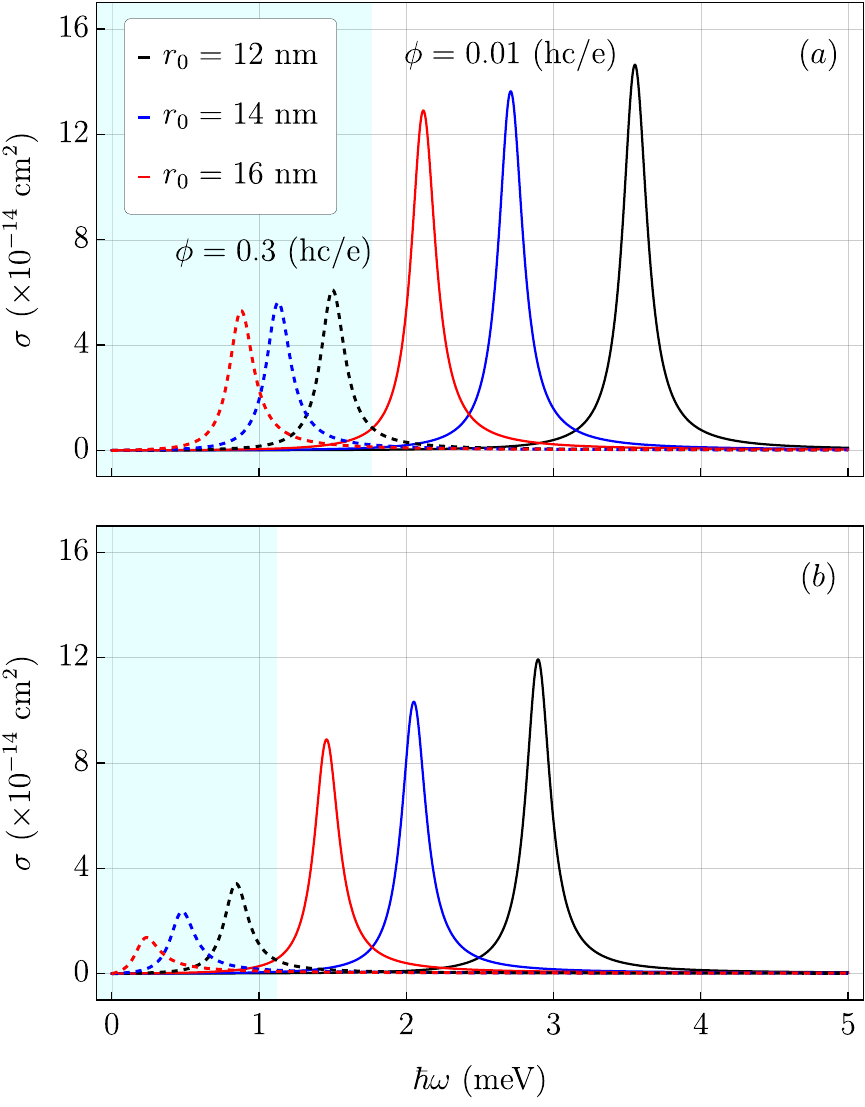}
    \caption{The same graphs as those in Fig. \ref{fig:flux}, representing the transition (${n=0, m=0}$) to (${n=0, m=1}$). (a) Corresponds to $\Omega= 0$, and (b) corresponds to $\Omega= 1\, \text{THz}$. The bluish region with solid curves corresponds to a fixed value of the magnetic flux, $\phi=0.3\, (hc/e)$, while the dashed curves outside this region represent the value of magnetic flux $\phi=0.01\, (hc/e)$.} 
    \label{fig:flux2}
\end{figure}

In contrast to the first transition, Fig. \ref{fig:flux2} illustrates that for the second transition (${n=0, m=0}$) to (${n=0, m=1}$), the effect is reversed, resulting in the peak position undergoing a shift towards lower energies as the magnetic flux increases. In this scenario, the amplitudes of the PCS peaks decrease when rotation is considered: the peaks in Fig. \ref{fig:flux2} (a) are higher than those in Fig. \ref{fig:flux2} (b). Additionally, the curves are closer to the origin compared to the previous case.
\begin{figure}[!t]
    \centering
    \includegraphics[scale=0.55]{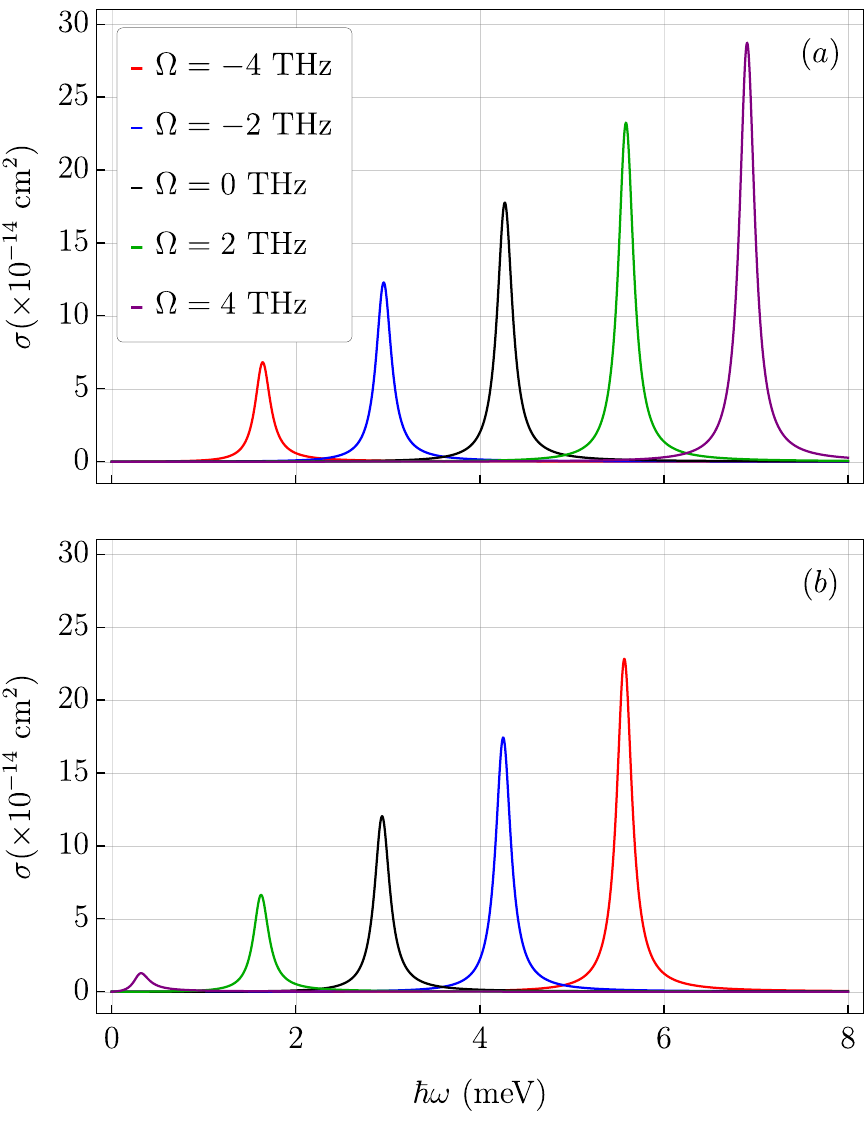}
    \caption{Graphs of the PCS for different values of the rotating parameter.  We have fixed the values of \(r_{0}=12\, \text{nm}\) and \(\hbar\omega_{0}=25\, \text{meV}\), magnetic flux $\phi=0.1\, (hc/e)$. (a) For the transition (${n=0,m=0}$) to (${n=0,m=-1}$) and (b) for the transition (${n=0,m=0}$) to (${n=0,m=1}$).} 
    \label{fig:rotation}
\end{figure}

Figure \ref{fig:rotation} presents the PCS for two distinct optical transitions as the rotation parameter varies, encompassing negative values. The first transition, from the state (${n=0, m=0}$) to (${n=0, m=-1}$) (Fig. \ref{fig:rotation} (a)), demonstrates an increase in amplitude with rotation. Moreover, there is a noticeable shift in the peak position towards higher values of photon energy.
As anticipated, the effect of rotation on the second transition (between the states (${n=0, m=0}$) and (${n=0, m=1}$) in Fig. \ref{fig:rotation} (b)) is opposite. The amplitude decreases with rotation, and there is a shift in the peak position towards lower values of photon energy. Consequently, we observe that the PCS curves for the second transition are more compact and positioned closer to the origin. Beyond that, the amplitudes of the peaks for the second transition are, in general, lower.

\section{Summary and conclusions}
\label{conc}

In this letter, we evaluated the PCS as a function of the energy of the incident photon for the two lowest optical transitions. We observed that the first transition ($n=0, m=0$) to ($n=0, m=-1$) exhibits a greater probability of realizing the photoionization process compared to the second transition ($n=0, m=0$) to ($n=0, m=1$). The peaks of PCS amplitude for the first transition are higher than those for the second transition. Moreover, as we increase the magnitude of the magnetic flux and rotation parameter, the PCS peaks also increase. Additionally, the energy of the incident photon increases with the consideration of these two parameters, leading to a larger range of possibilities for the release of the bound electron.

In conclusion, we would like to mention that we are currently investigating novel effects, such as the incorporation of spin effects on the optical properties (such as the photoionization process) of some low-dimensional quantum systems. We are also exploring new methods aimed at solving their wave equations. Any significant advancements resulting from this research will be reported in future publications.

\section*{Acknowledgments}

This work was partially supported by the Brazilian agencies CAPES, CNPq, and FAPEMA. E. O. Silva acknowledges CNPq Grant 306308/2022-3, FAPEMA Grants PRONEM-01852/14 and UNIVERSAL-06395/22. F. S. Azevedo acknowledges CNPq Grant No. 150289/2022-7. This study was financed in part by the Coordena\c{c}\~{a}o de Aperfei\c{c}oamento de Pessoal de N\'{\i}vel Superior - Brasil (CAPES) - Finance Code 001.

\bibliographystyle{apsrev4-2}
%

\end{document}